\documentclass[%
 amsmath,amssymb,
 aps,
 prfluids,
 onecolumn, 
 11pt
]{revtex4}

\usepackage{graphicx}
\usepackage{dcolumn}
\usepackage{bm}
\usepackage{xcolor}
\usepackage{svg} 
\usepackage{amsmath} 
\usepackage[normalem]{ulem}
\usepackage{txfonts}

\newcommand{\REV}[1]{\textcolor{black}{#1}} 


\begin{document}

\preprint{APS/123-QED}

\title{Dissociation of red blood cell aggregates in extensional flow}

\author{Midhun Puthumana Melepattu}
  \email{Midhun.Puthumana-Melepattu@univ-grenoble-alpes.fr}
\author{Guillaume Ma\^{\i}trejean} 
\email{Guillaume.Maitrejean@univ-grenoble-alpes.fr}
\author{Thomas Podgorski}%
 \email{Thomas.Podgorski@univ-grenoble-alpes.fr}
\affiliation{
Université Grenoble Alpes, CNRS, Grenoble INP, LRP, 38000 Grenoble, France}
\altaffiliation[]{The authors are members of LabEx Tec21 (Investissements d'Avenir - grant agreement ANR-11-LABX-0030)}

\date{July 14, 2024}

\begin{abstract}

Blood rheology and microcirculation are strongly influenced by red blood cell aggregation. We investigate the dissociation rates of  red cell aggregates in extensional flow using hyperbolic microfluidic constrictions and image analysis by a convolutional neural network (CNN). Our findings reveal that aggregate dissociation  increases sharply when a critical extension rate is reached which falls within the range of microcirculatory conditions, suggesting that large variations of aggregate sizes should be expected in-vivo. This work contributes to a deeper understanding of the behavior of red blood cell aggregates in response to extensional stress in microcirculatory networks, provides crucial experimental data to validate theoretical and numerical models, and constitutes the basis for improved evaluation of  blood  aggregability in clinical contexts.

\end{abstract}

\keywords{Blood, aggregation, red blood cell, microfluidics, artificial intelligence, machine learning, CNN}
\maketitle

Plasma proteins, mainly fibrinogen, are responsible for a reversible aggregation of red blood cells (RBCs) \cite{baskurt2011}: near stasis and under low hydrodynamic stresses, the flat equilibrium shape of RBCs promotes their aggregation into clusters called ‘rouleaux’ that resemble stack of coins, eventually interconnecting to form 3D networks. 
As hydrodynamic stresses rise and overcome aggregation forces, the disaggregation of these structures, along with RBC deformability, results in the well-known shear-thinning behavior of blood \cite{chien1970}. 
Aggregation directly affects blood sedimentation, forming the basis of the erythrocyte sedimentation rate (ESR) test, a non-specific inflammation marker \cite{bedell1985}.

Two mechanisms of aggregation have been identified. A bridging mechanism was first proposed, supported by the evidence of macromolecule adsorption on the surface of RBCs and the role of electrostatic forces \cite{Chien1973RedCA} to explain the non-monotonous nature of the interaction force as a function of the macromolecular concentration. Both specific \cite{bagchi2005} and non-specific \cite{lominadze2002,perevezev2005} ligand interaction models have emerged. A second mechanism is based on depletion effects \cite{asakura1958,neu2002} due to the finite size of macromolecular aggregation promoters such as fibrinogen or Dextran. While evidence shows that both mechanisms coexist in experimental situations, there are still open questions regarding their relative weight, depending on conditions (concentrations, synergies between fibrinogen and other factors, Dextran of model experiments,  alterations of the RBC membrane, etc.).

While shear rates in blood circulation are usually above 100 s$^{-1}$ \cite{Robertson2008,Podgorski2022}, a range where no influence of aggregation is detected in macroscopic rheology measurements \cite{chien1970}, several works have revealed that aggregation interactions determine the structure of RBC suspensions in small channels \cite{Bishop2001,Zhang2009}, stabilize clusters in capillaries \cite{brust14} or influence blood perfusion in networks \cite{Reinhart2017}. A prominent feature of blood microcirculation is the heterogeneity of hematocrit distribution,
a consequence of the Zweifach-Fung effect leading to an uneven distribution of RBCs at bifurcations, generally resulting in an increased hematocrit in the downstream branch with the highest flow  \cite{fenton85,doyeux11,Shen16}.
While studies have focused on the influence of aggregation on hematocrit distribution, 
albeit in relatively wide channels 
\cite{Sherwood2014,kaliviotis17}, the stability of RBC aggregates in bifurcations has seldom been explored \cite{yayathesis}
although it is a strong determinant of aggregate size distribution in capillary networks.
Indeed, aggregates evolve only slowly in straight channels or capillaries \cite{brust14,Claveria16} and the distance between bifurcations is short in-vivo \cite{Pries1995}. On the other hand, significant extensional stresses in bifurcations of the microcirculation may control aggregate dissociation and size distribution.

Several studies, mostly computational, investigated the aggregation and dissociation dynamics of RBCs in shear flows \cite{bagchi2005,Zhang2008} and associated rheological consequences \cite{Fedosov2011}. This configuration is relevant to blood dynamics in large vessels down to arterioles or venules but not so much to circulation in capillaries where the confinement of RBCs and clusters is such that shear does not significantly affect aggregates. In contrast, their behavior in generic extensional flows, relevant in microcirculation scenarios with significant extensional stresses at bifurcations, has been rarely studied. Indeed, while bifurcations vary a lot in vivo (bifurcation angle, relative diameters and flow rates in branches), they share a localized strong extensional component of the stress tensor, prompting a need for understanding aggregate behavior in such conditions. In addition, extensional flows offer an alternative to costly and time consuming techniques like AFM \cite{steffen2013quantification} for quantifying  aggregation forces or energy at the cellular level.

In this work, we provide the first quantitative experimental study of RBC disaggregation in extensional flows using microfluidic hyperbolic constrictions.  We focus on the impact of aggregation strength and flow extension stress on dissociation probabilities, proposing a straightforward scaling law. Our findings enhance comprehension of red blood cell aggregate behavior in microcirculatory networks, offering crucial experimental data for numerical model comparison and benchmarking. 
The proposed experimental principle, which can readily be used to study aggregate dissociation with RBCs in their own plasma rather than in Dextran solutions, provides a high-throughput detailed measurement of the distribution of aggregate stability in a blood sample and an indirect measurement of interaction strength thanks to the proposed scaling law. This could form the basis of a RBC aggregability measurement that would complement existing techniques.

\begin{figure}[t!]
    \centering
   \includegraphics[width = 0.6\linewidth, ]
   {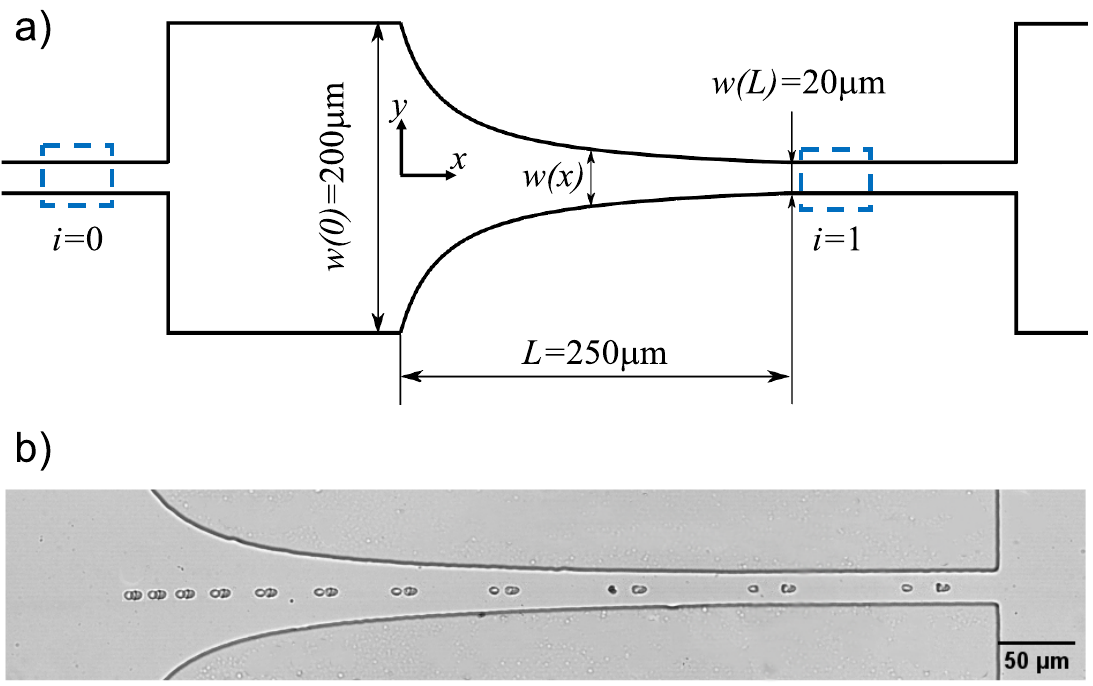}
    \caption{Hyperbolic flow geometry producing extensional flow: (a) Design of the central part of the microfluidic channel and notations. Blue areas: image processing ROIs;   (b) 
    Example temporal sequence of a 3-cell aggregate breaking into $2+1$ cells in the extensional flow with $\dot{\varepsilon}=$ 204 s$^{-1}$ (Time step: 4 ms; Dextran concentration 3 g/dL). 
    }
    \label{fig:constriction}
\end{figure}

 \textit{Experiment -} The central part of the microfluidics setup is a hyperbolic constriction (Fig. \ref{fig:constriction}a) producing a nearly planar extensional flow thanks to a decreasing channel width  
$w(x)\sim 1/x$, where $x$ is the  coordinate in the flow direction,
 as implemented in other works \cite{bento2018deformation,faustino2019microfluidic}.
Results presented here were obtained using $w(0)=200~\mu$m, $w(L)=20~\mu$m and a thickness $h=w(L)$ in the $z$ direction and $L=250~\mu$m.
 Mass conservation implies that the cross-sectional average of the axial velocity $<u_x(x)>$ increases linearly along $x$ i.e. the average extension rate $d<u_x(x)>/dx$ is constant.
 Given that the Reynolds number $Re=\rho v(L) h/ 2\eta$ (where $\rho$ and $\eta$ are the density and viscosity of the suspending fluid) is below 1 in all experiments and assuming that the velocity along the central axis $v_x(x)=u(x,0,0)$ increases linearly  as the average velocity, a fluid particle flowing along the centerline will experience a nearly constant fluid extension rate defined as $\dot{\varepsilon_f} = {d v_x}/{dx}$. COMSOL numerical simulations 
 were performed to confirm the quasi-linearity of the velocity profile
 (see Supplemental Material \cite{suppmat}).
 
In practice, the channel features a succession of constrictions with fixed inlet and outlet widths $w(0)=200~\mu$m, $w(L)=20~\mu$m, and decreasing lengths $L=1000, 500, 250, 125 ~\mu$m exposing clusters to increasing extension rates. Each of them is followed by a straight channel (cross section $w(L)\times h=20 \times 20~\mu$m$^2$, length 150$\mu$m) in order to facilitate image acquisition at the end of the extensional flow region, and a large channel section ($w=200 ~\mu$m, $L=150 ~\mu$m) ensuring that clusters are relaxed before entering the next constriction.
Note that due to strong confinement in the $z$ direction, the average shear rate is actually quite high and of order $\dot{\gamma}\sim 2 v_x(x)/h$, meaning that $\dot{\gamma}/\dot{\varepsilon_f} \sim 25$ at the end of the constriction. 
However, {as already seen in previous works \cite{brust14,Claveria16},} in this   
configuration where flowing RBC aggregates are well centered in the channel the symmetric shear forces do not significantly contribute to disaggregation.
Therefore, we expect extensional stresses to provide the dominant dissociation mechanism{, which is confirmed by our observations below}.

Blood samples  {provided through an Agreement for the Transfer of Non-Therapeutic Blood Products with the French Blood Establishment (EFS) (Ref. EFS AURA 22-082)} were collected from healthy donors and anticoagulated with citrate, stored at 4°C and used within 1 week after collection for consistent results \cite{Merlo2023}.
{A total of 7 samples from different donors were used in the presented results.}
Before experiments, RBCs were washed by successive dilutions with Phosphate Buffered Saline (PBS, Sigma-Aldrich, P4417) and centrifugations (1000g $\times$ 3 min, 3 times).
Aggregation  was  controlled by resuspending RBCs at an initial hematocrit of 2--3 $\%$ in PBS supplemented with Dextran 70 kDa (Sigma-Aldrich, D8821), as in previous studies \cite{brust14,flormann2017}. 
Dextran, promotes aggregation by favoring mainly depletion interactions \cite{brust14,flormann2016,flormann2017,yaya2021,steffen2013quantification} with a possible contribution of bridging interactions \cite{yayathesis}. By varying the Dextran concentration from 1 to 5 g/dL, the interaction energy varies in the range $[-5, -1]$ $\mu$J/m$^2$, as measured by AFM  \cite{steffen2013quantification,brust14}. 
RBCs were introduced in the channel through a flow focusing device  that allows to dilute RBCs with the PBS-Dextran solution and focus aggregates around the centerline before they reach the constriction section {where further centering is achieved after the first constriction - see Supplemental Material \cite{suppmat} for a quantification}. The flow was controlled by an Elveflow OB1 pressure controller, allowing to vary extension rates in constrictions between 30-650 $s^{-1}$. {The slow velocity of cells in the inlet section ahead of the constrictions allowed for a visual check to ensure that cells had a healthy discocyte shape when entering the channel. The channel had also initially been passivated with a solution of 1 g/L Bovine Serum Albumin (BSA, Sigma-Aldrich, A7906) to prevent undesirable interactions between cells and walls.}

Observations were made 
in bright field microscopy with an Olympus IX71  microscope equipped with a x10 objective lens and a Photron Fastcam Mini-UX100 fast camera (resolution 1.4 pixel/$\mu$m).  A  typical image sequence showing aggregate dissociation is shown in Fig. \ref{fig:constriction}b.
In order to evaluate the actual or effective particle extension rate experienced by aggregates $\dot{\varepsilon}=dv_p/dx$
where $v_p$ is the particle velocity, sequential positions $x(t)$ of a few particles (aggregates) along the hyperbolic constriction were measured and fitted with a function $x(t)=a+b \exp(\dot{\varepsilon} t)$.
Note that each aggregate passes through and survives a straight channel with a cross-section of $20 \mu$m $\times$ $20 \mu$m (where the shear exerted is maximum) before entering the hyperbolic constriction. Hence, any breaking events to the clusters passing through the hyperbolic section must be triggered by the extension rather than shear.
This is confirmed by our observation that all breaking events take place significantly before the end of the constriction ($x<150~\mu$m), i.e. in a region where shear is much lower compared to the straight section following the previous constriction. Since the only significant difference, in terms of hydrodynamic stresses, between the studied constriction and the straight $20 \mu$m wide channel located upstream is the additional extensional component, one can safely conclude that aggregate dissociation is controlled by extensional stress in our device and that $\dot{\varepsilon}$ is the relevant parameter.

The evolution of RBC cluster size distribution was evaluated using successively classical image processing and a custom convolutional neural network (CNN) algorithm  (see Supplemental Material \cite{suppmat}). 
 Image processing involves selecting two regions of interest (ROI) in  straight 20~$\mu$m-wide sections of the channel located before and after a given constriction in order to compare aggregate size distributions.   
We verified that no re-aggregation occurs in the abruptly expanding zones after  constrictions thanks to the very short residence time of cells.
 Datasets consisting of images of the two ROIs are then processed by the CNN for classification.

\begin{figure}[t!]
     \centering
 \includegraphics[width = 0.65\linewidth, ]{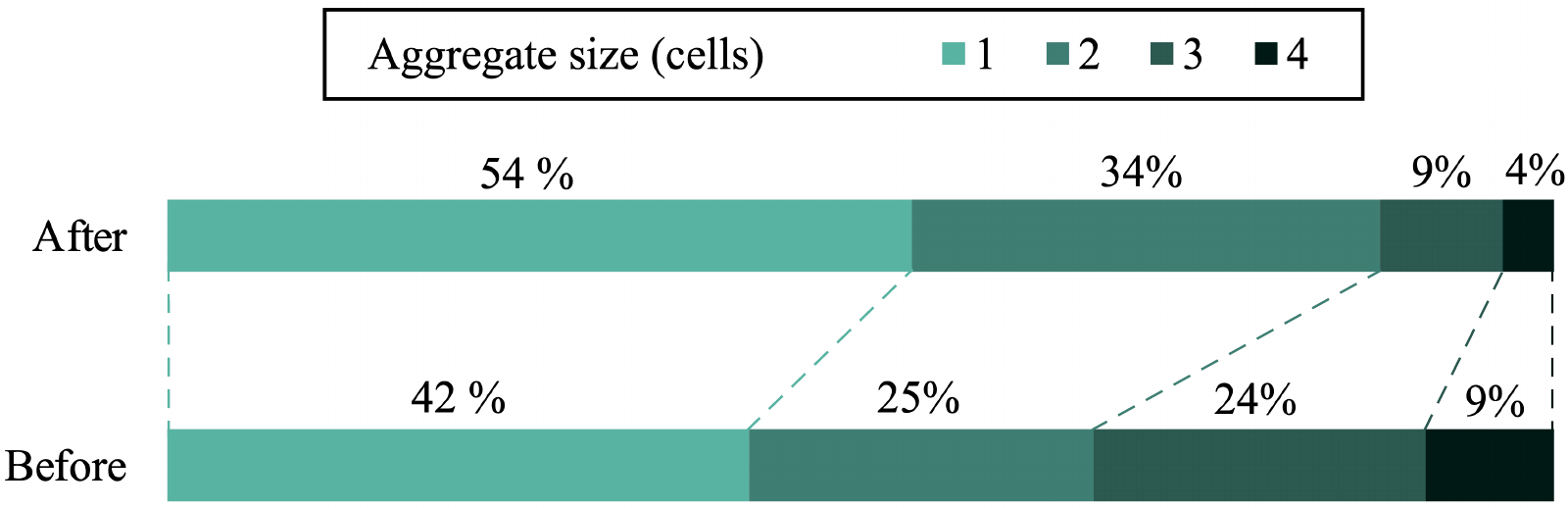}
   \caption{Distribution of the RBC population in aggregates of different sizes before and after a constriction, for an extension rate $\dot{\varepsilon}=204$ s$^{-1}$ and a Dextran concentration of 3 g/dL.
   }
     \label{fig:bargraph}
 \end{figure}

 Overall, the image processing allows to detect and analyze between 1000 and 10000 objects for a given experiment sequence (fixed flow rate and Dextran concentration), which are then classified into single cells, doublets, triplets, quadruplets, and larger objects with respective relative populations $S_i$,  $D_i$,  $T_i$,  $Q_i$ and $N_i$ (with $i=0$ before and $i=1$ see Fig. \ref{fig:constriction}a). These numbers are normalized by the total number of detected RBCs (aggregated or not), such that $S_i+2 D_i+3 T_i + 4 Q_i=1$ (aggregates of 5 cells or more are very rare and not included in this normalization). An example of the evolution of the distribution of RBCs in aggregates of different sizes (i.e. relative values of $S_i$, $2 D_i$, $3 T_i$ and $4 Q_i$) when going through the constriction is shown in Fig. \ref{fig:bargraph}, reflecting the disaggregation of large aggregates in to smaller ones and single cells.

We then define dissociation the probabilities $p_d$ (for doublets),  $p_t$ (triplet breaking into a doublet and a single cell), $p_{q1}$ and $p_{q2}$ (quadruplet breaking into respectively two doublets or a single cell and a triplet), with $p_q=p_{q1}+p_{q2}$.
Note that based on observation the
probability of an aggregate breaking into more than two parts was considered negligible, likely due to stress relaxation when a first bond breaks. 

Assuming mass conservation yields the following relations:
  \begin{subequations}
    \begin{eqnarray}
     S_1 = S_0+ 2p_d D_0+ p_t T_0 + p_{q1} Q_0 \label{eqa}
    \\
     D_1 = D_0 + p_t T_0 + 2 p_{q2} Q_0 - p_d D_0 \label{eqb}
     \\
     T_1 = T_0 + p_{q1} Q_0 - p_t T_0 \label{eqc}
     \\
     Q_1 = Q_0 (1-p_{q1}-p_{q2})  \label{eqd}
    \end{eqnarray}
    \end{subequations}

\begin{figure*}[t!]
 \resizebox{0.95\textwidth}{!}{
   \includegraphics{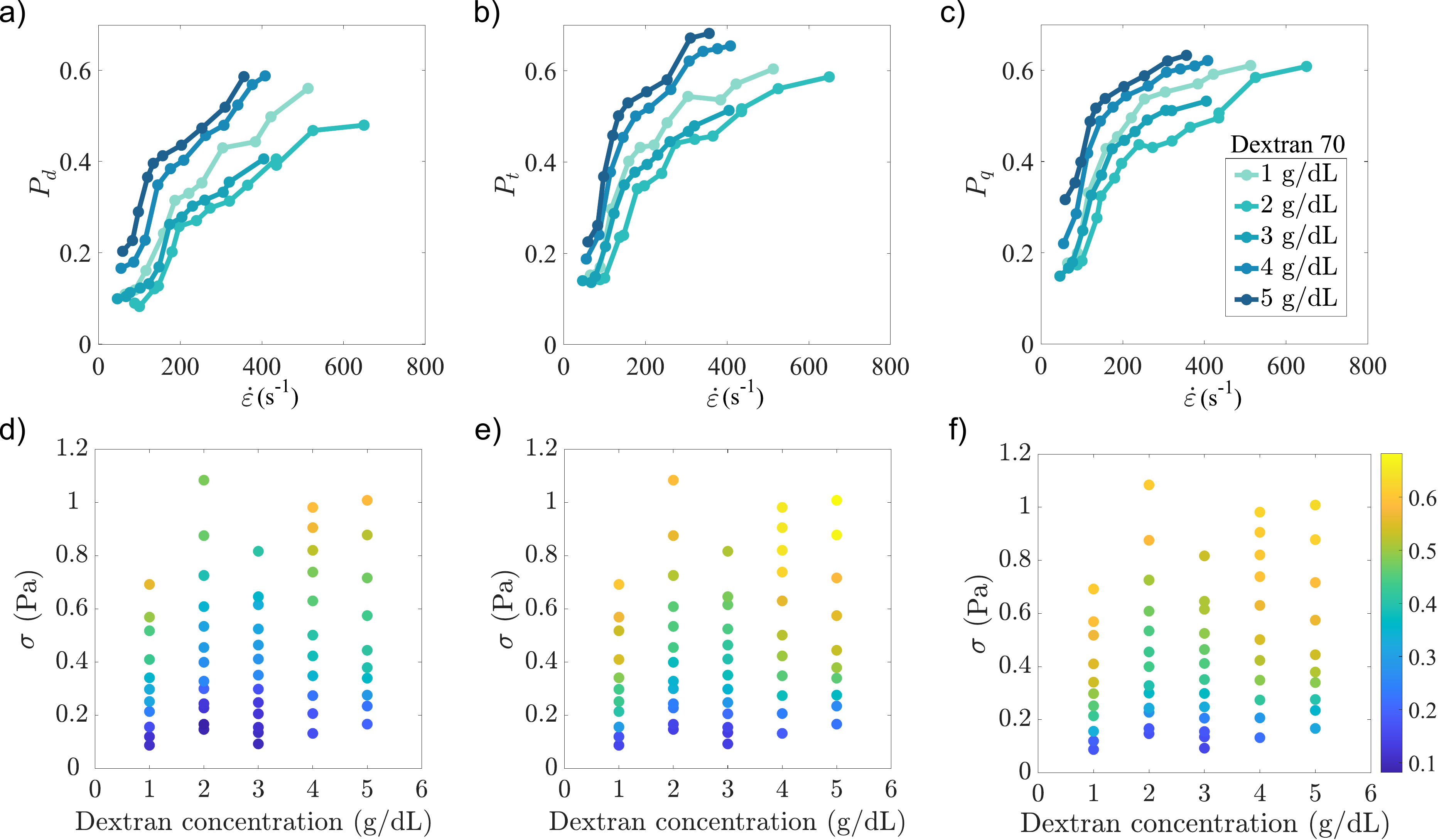}}
  \caption{
  Aggregate dissociation probabilities as a function of extension rate and Dextran 70 kDa concentration. (a), (b), (c): disaggregation probabilities of doublets ($p_d$), triplets  ($p_t$) and quadruplets ($p_q$) vs. extension rate $\dot{\varepsilon}$ for different Dextran concentrations, showing a sharp transition between 80 and 200~s$^{-1}$; (d), (e), (f): disaggregation diagrams in the extension stress $\eta \dot{\varepsilon}$ vs. Dextran concentration parameter space, respectively for doublets, triplets, and quadruplets. The color scale represents dissociation probabilities. 
  \label{fig:breakingstats}}
\end{figure*}

This set of linear equations, which is redundant due to mass conservation can be solved to derive  disaggregation probabilities $p_d$, $p_t$ and $p_q=p_{q1}+p_{q2}$ by making the additional assumption that the ratio $\lambda =p_{q2}/p_{q1} $ is constant. A sensitivity analysis reveals that the solution of the system of equations only weakly depends on the exact value of $\lambda$. Therefore, we made the reasonable assumption that the three different cell-cell bonds in a linear quadruplet had an equal probability of breaking, that is, $\lambda=1/2$. The results are summarized in Fig. \ref{fig:breakingstats} where these probabilities are plotted as a function of extension rate and Dextran concentration (See Fig. S5 of the Supplemental Material \cite{suppmat} for comments on sample variability.).

Remarkably, dissociation probabilities show a sigmoid behavior when increasing the extension rate with a strong jump in the range $80-200$~s$^{-1}$ that hints at a critical extension rate $\dot{\varepsilon}_c$ and a slower increase at higher extension rates.
Ideally, if all aggregates were identical, the same reproducible scenario would apply and they would break at the same extension rate, when flow stresses overcome interaction forces.
The dissociation probabilities of Figs. \ref{fig:breakingstats}(a,b,c) would then be step functions with zero breaking probability below $\dot{\varepsilon}_c$ and 1 above.

The smoother progression of dissociation probabilities is likely due to several factors. Firstly, the dispersion in orientations and morphologies of aggregates entering the constriction causes variations in aggregate strength and stress. This phenomenon, noted in early studies of aggregate dissociation in shear flow \cite{Skalak1984},, affects dissociation probability. Secondly, the heterogeneity of cell properties is significant: RBC aging increases stiffness \cite{Waugh1992Mar,bosch1994,franco2013}, impacting aggregation properties \cite{hoore2018}.

After the transition in the range $80<\dot{\varepsilon} <200$s$^{-1}$, the dissociation probability increases slowly and neither reaches 1 nor shows asymptotic behavior at the highest rates tested. Doublets, with an aspect ratio close to 1 and less prone to alignment in the flow direction, are particularly robust, with less than 50\% dissociating at the highest rate for Dextran concentrations of 2-3 g/dl. Although aggregates would eventually break under sufficiently high extension rates and prolonged time, this experimental system cannot vary these factors independently. In the microcirculation, RBC aggregates mainly encounter extensional stresses in bifurcating vessels, making brief periods of extension physiologically significant.

Qualitatively, larger aggregates (Figs. \ref{fig:breakingstats}(b,c)) tend to break at slightly lower extension rates than smaller ones, with significantly higher dissociation rates. This is consistent with the higher total drag force exerted on the surface of longer aggregates, which has to be balanced by attraction force at the weakest cell-cell contact zone in the aggregate.

Figs. \ref{fig:breakingstats}(d-f) display disaggregation diagrams: To account for the non-negligible variation of suspending medium viscosity $\eta$ with Dextran concentration $c$, we represent disaggregation probabilities in the $(\sigma,c)$ parameter space 
where $\sigma= \eta \dot{\varepsilon}$ is the extensional stress. Following previous measurements \cite{flormann2016}, we take $\eta(c)=-1.78+2.84 \exp(0.0097 c)$ mPa.s with $c$ in mg/ml. Interestingly, the dissociation probabilities 
defined above follow a marked non-monotonous bell-shaped behavior as in several other RBC aggregation studies with Dextran as an aggregation promoter \cite{neu2002,brust14,flormann2016,flormann2017}. 
The comparison of data for
doublets, triplets, and quadruplets  reveals a consistent pattern wherein the maximum aggregate strength is achieved at Dextran concentrations ranging between 2 and 3 g/dl with, again, a global shift towards lower stresses that indicates a higher fragility of larger aggregates.

Assuming a balance between the work of hydrodynamic forces and the interaction energy at the contact zone of two cells, the critical dissociation stress can be estimated through dimensional arguments. For two rounded cells of radius $R$ aligned with the flow, the contact area is a disk with radius $R$ (area $\sim \pi R^2$). In the moving frame at the doublet's center, it is in a planar elongational flow with extension rate $\dot{\varepsilon}$ and $x$-velocity $\dot{\varepsilon}x$. The average relative fluid velocity with respect to the right cell is $U \sim \dot{\varepsilon} R$ (and $-U$ on the left side), so each cell experiences a drag force similar to the Stokes force on a sphere of radius $R$.
\begin{equation}
F \sim \pm 6 \pi \eta R U \sim  \pm  6 \pi \eta \dot{\varepsilon} R^2 
\end{equation}
with a positive sign for the right cell and a negative sign for the left cell. These two opposing forces tend to separate the two cells in the moving frame.
Assuming cells are pulled apart a distance $\delta l$  from the origin of the moving frame, these forces produce a total work: 
\begin{equation}
W= 2 * 6 \pi \eta \dot{\varepsilon} R^2 \delta l
\label{work}
\end{equation}

This work compensates the loss of interaction energy between cells. Denoting $\epsilon_{ad}$ the interaction energy per unit area as defined earlier \cite{steffen2013quantification,neu2002, brust14, flormann2017}, the total adhesion energy is:
\begin{equation}
E_{ad}= \epsilon_{ad} * \pi R^2
\label{adenergy}
\end{equation}

By equating $W=E_{ad}$ one gets the critical extension rate required to overcome adhesion:
\begin{equation}
\dot{\varepsilon} _c = \epsilon_{ad} / (12 \eta \delta l)
\label{critepsilon}
\end{equation}

Note that in this simple energy balance, we neglected viscous dissipation due to the peeling of RBC membranes from each other. This assumption is based on the large contact angle between membranes resulting from their deformability (see the first snapshots in Fig. \ref{fig:constriction}b) and is supported by  features of the experimental results: If viscous dissipation between separating membranes were the major force opposing dissociation, the results would be nearly independent of viscosity, extension rate, and interaction energy \REV{(See also Supplemental Material \cite{suppmat} for an estimation of the lubrication-mediated separation timescale).}

Taking $\epsilon_{ad}=5$~$\mu$J/m$^2$ (maximum value of the interaction energy in Dextran 70 kDa solution as measured in Steffen et al\cite{steffen2013quantification}), $\eta = 2$mPa.s (approximate viscosity of 3g/dL Dextran solution) and
$\delta l \sim 1$~$\mu$m (order of magnitude of the displacement required to separate cells), equation \ref{critepsilon} yields a critical extension rate of order 200 s$^{-1}$ which is the order of magnitude we measured in experiments (Fig. \ref{fig:breakingstats}a).

For larger aggregates (triplets, quadruplets), as the centers of mass of the leftmost and rightmost cells are at a larger distance from the origin of the moving frame, a smaller extension rate is needed to produce the same relative fluid velocity and drag force. Using the same scaling argument, the extension rate required to dissociate an aggregate of $N$ cells becomes 
\begin{equation}
\dot{\varepsilon} _c(N) = \dot{\varepsilon}_c/(N/2)= \epsilon_{ad} / (6 N  \eta \delta l)
\label{critepsilonN}
\end{equation}

\begin{figure}
    \centering
 \includegraphics[width = 0.9\linewidth ]{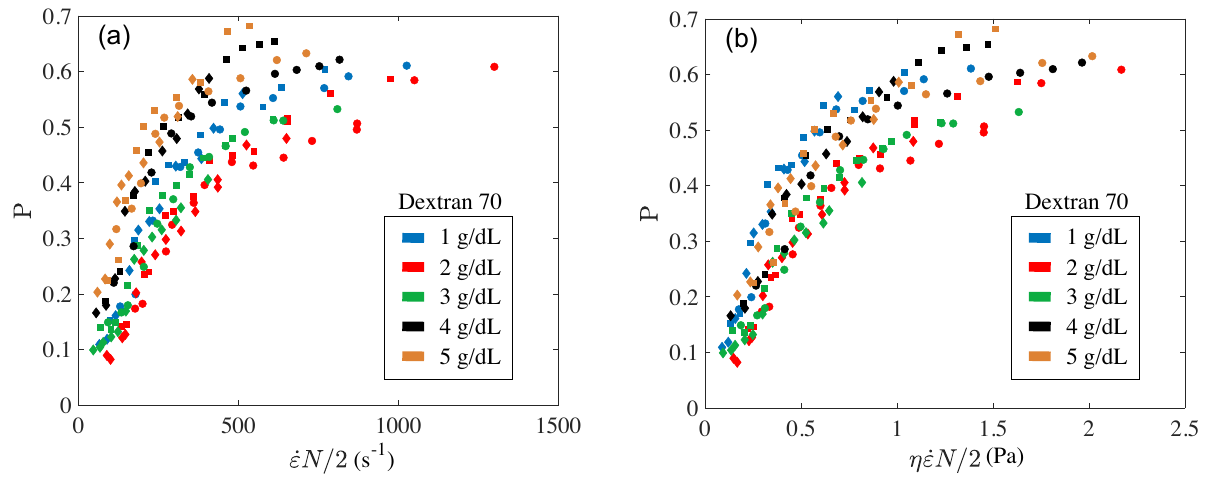}
    \caption{Aggregate dissociation probability vs. $\dot{\varepsilon}N/2$ (a) and vs. $\eta \dot{\varepsilon} N/2 $ (b) as suggested by Eq. \ref{critepsilonN} 
    showing a collapse of data for all aggregate sizes ($\vardiamondsuit$: doublets, $\blacksquare$: triplets, $\medbullet$: quadruplets) and an ordering from left to right consistent with increasing interaction energy $\epsilon_{ad}$.
    }
    \label{fig:stress_scaling}
\end{figure}

This scaling is tested plotting the dissociation probabilities as a function of $\dot{\varepsilon}(N/2)$ (Fig. \ref{fig:stress_scaling}a) and $\eta \dot{\varepsilon}_c N/2 $ (Fig. \ref{fig:stress_scaling}b), the only independent parameter left being the interaction energy $\epsilon_{ad}$. 
A nice collapse of experimental data corresponding to different aggregate sizes is obtained \REV{in Fig. \ref{fig:stress_scaling}a, especially considering the initial scattering of individual data series in Fig. \ref{fig:breakingstats}a-c}, as well as an ordering according to interaction energy in Fig. \ref{fig:stress_scaling}b with weak aggregation cases on the left side (1, 4 and 5 g/dL) and \REV{ a nearly perfect overlap of data for 2 and 3 g/dL consistent with similar values of the adhesion energy near the maximum of the bell-shaped adhesion energy curve \cite{brust14,steffen2013quantification}. }

Interestingly, the Dextran 70kDa concentration at which the maximum influence of aggregation forces is observed appears highly dependent on experimental setup and  stress type: while here we report a maximum resistance to extension at 2-3 g/dL as in  AFM measurements \cite{steffen2013quantification} or for clusters in straight microchannels \cite{brust14}, other studies reported a peak of aggregation-related phenomena in the range 4-6 g/dL \cite{flormann2016,flormann2017}. This variation may stem from differences in cell-cell interactions across various flow conditions, potentially leading to shifts in Dextran concentration influence.

In conclusion, the response of RBC aggregates to extensional flow stresses was quantitatively investigated by studying dissociation statistics under various levels of stress and interaction forces. This provides an original dataset that should be useful for the validation of RBC aggregation models and the benchmarking of numerical simulations.
Remarkably, strong changes in behavior are found in a range of parameters  corresponding to physiological situations: (i) the range of extension rates, around 100 s$^{-1}$, where transitions are seen corresponds to typical extension rates found at bifurcations of the microcirculation in-vivo (velocities around 1 mm/s in capillaries with diameters around 10 $\mu$m) and (ii) considering previously measured values of the interaction energy \cite{steffen2013quantification,brust14}, the normal range of physiological fibrinogen concentrations (1.8-4 g/L) corresponds to approximately 1 g/dL of Dextran 70kDa while the concentrations at which we see the most robust aggregates (2-3 g/dL of Dextran) correspond to elevated fibrinogen levels seen in pregnancy, inflammatory and other hyper-aggregability cases. This suggests that in-vivo variations of fibrinogen levels between normal and pathological cases could lead to significant differences in dissociation rates of RBC aggregates in  microcirculatory networks and with a likely strong impact on the distribution of aggregate sizes. This is expected to significantly influence the local rheology of blood and hematocrit distribution in capillary networks and calls for further investigations on the behavior of aggregates in complex networks and their impact on blood perfusion.

Finally, the experimental and data analysis principle  developed here offer a alternative and relevant way to characterize RBC aggregation that could be used for fundamental studies on the influence of RBC mechanical properties, aggregation mechanisms (using different aggregation promoters such as fibrinogen or pure depletant agents \cite{Korculanin2021}) or for an improved evaluation of the aggregability of healthy and pathological blood samples in clinical contexts.

\textit{Acknowledgements.}
We thank M. Van Melle-Gateau (LIPhy, CNRS-UGA) for microfabrication, J. Martin-Wortham for experimental advice and M. Karrouch for technical support. T.P. thanks G. Ghigliotti, G. Coupier, F. Yaya, C. Boucly and C. Wagner for discussions. This work was supported by CNES (Centre National d'Etudes Spatiales, DAR ID 8106 "Rhéologie sanguine").

\bibliographystyle{unsrt}

\bibliography{Disaggregation}

\providecommand{\noopsort}[1]{}\providecommand{\singleletter}[1]{#1}%

\end{document}


\preprint{APS/123-QED}

\title{Supplemental Material: Dissociation of red blood cell aggregates in extensional flow}

\author{Midhun Puthumana Melepattu}

  \email{Midhun.Puthumana-Melepattu@univ-grenoble-alpes.fr}
\author{Guillaume Ma\^{\i}trejean} 
\email{Guillaume.Maitrejean@univ-grenoble-alpes.fr}
\author{Thomas Podgorski}%
 \email{Thomas.Podgorski@univ-grenoble-alpes.fr}
\affiliation{
Université Grenoble Alpes, CNRS, Grenoble INP, LRP, 38000 Grenoble, France
}%

\date{July 11, 2024}


\maketitle

\section{COMSOL Simulations}

\begin{figure}[ht!]
\centering
 \includegraphics[width = 0.95\linewidth]{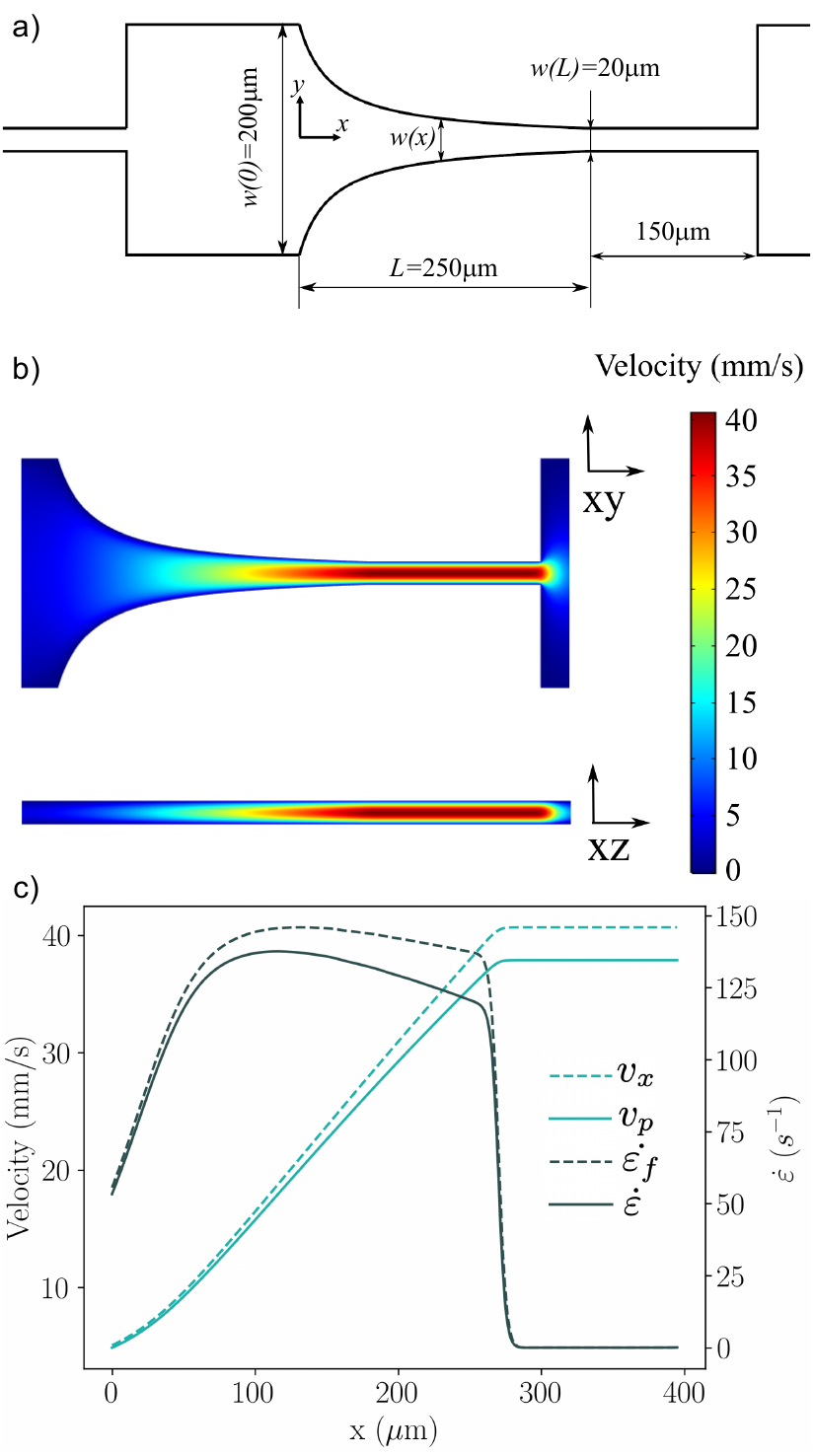}
  \caption{Hyperbolic flow geometry: (a) Design of the simulated flow geometry (250 $\mu$m long constriction (thickness $h=20$ $\mu$m, $w(0)=200 \mu$m, $w(L)=200 \mu$m) followed by a 150 $\mu$m long straight section); (b) Result of COMSOL flow simulation with parameters: viscosity $\eta=1$ mPa.s, density $\rho=1$ g/cm$^3$, inlet pressure $P_{in}=400$ Pa, outlet pressure $P_{out}=0$ Pa.  The color scale codes for the norm of the axial flow velocity $u_x(x,y,z)$ in median cross-sections in the $(0xy)$ and $(0xz)$ planes; (c) Velocities (green) and extension rates (black) along the channel from the COMSOL simulation shown in (b). Values along the central axis $v_x(x)=u_x(x,0,0)$ and $\varepsilon_f=dv_x/dx$ are in dashed lines while particle velocity $v_p(x)$ and the corresponding extension rate $\varepsilon=dv_p/dx$ (full lines) are estimated from the average of $u_x(x,y,z)$ over $y \in [-3.5 \mu$m$,3.5\mu$m$]$ and $z \in [-3.5 \mu$m$,3.5\mu$m$]$. 
  \label{fig:constriction}} 
\end{figure}

In order to confirm that a constant extension rate will actually be produced along the centerline in a 3D hyperbolic constriction with a finite thickness $h$ and quantify the effective extension rate applied to particles with a finite size, we performed numerical simulations of the Navier-Stokes equation using COMSOL Multiphysics in the channel geometry depicted in Fig. \ref{fig:constriction}(a). 
 Varying the thickness $h$ and final width $w(L)$ at the exit of the constriction between  10 $\mu$m and 40 $\mu$m, the evolution of the centerline velocity $v_x(x)$ was studied, as well as the approximate particle velocity $v_p(x)$ estimated by averaging the velocity field over an area of $7 \times 7$~$\mu$m$^2$ around the centerline.
 
 This led to an experimental design with final section $h=w(L)=20$~$\mu$m and entrance width $w(0)=200$~$\mu$m, as shown in Fig. \ref{fig:constriction}(a). 
 This final choice was motivated by a compromise between the possibility to reach high extension rates (requiring small $w(L)$ and a high ratio $w(0)/w(L)$) while avoiding too much interaction between cells and walls (requiring $h$ and $w(L)$ to be somewhat larger than RBC aggregate sizes). On the other hand it is necessary to  achieve a good centering of cells in the $z$ direction thanks to migration forces, requiring a rather low $h$ \cite{losserand19}.

 Simulation results using the final channel geometry (Fig. \ref{fig:constriction}b-c) show that after $x \simeq 50$~$\mu$m the central velocity and the particle average velocity increase nearly linearly with distance, allowing to define a particle extension rate $\dot{\varepsilon}=d v_p/dx$ that is approximately constant within $\pm 8$\%. The weak nonlinearity of $v_x$ and $v_p$ can be attributed to entrance effects at non-zero Reynolds number, the quick reduction of the cross-section in the first half of the channel, as well as an evolution of the details of the velocity profile as the cross-section goes from flat-rectangular to square. The evolution of the ratio of particle size to channel width also contributes to a small reduction of particle mobility in the second half of the channel\cite{vitkova04,hetsroni1970}. However, the $\pm 8$\% variation of particle extension rate can safely be considered acceptable compared to other sources of fluctuation in experiments (particle orientation and position in the channel, velocity measurements etc...).

\medskip

\medskip

\section{Focusing of cells}

The first part of the microfluidic chip is a generic 3-inlet cross-flow type device in which the dispersed phase (concentrated RBC suspension) is focused in the $y$-direction between two perpendicular inlets of the suspending phase, followed by a $\sim 5$ cm long channel that allows RBC aggregates to relax around the center plane in the $z$-direction thanks to migration forces \cite{losserand19}. All channels in this device have a rectangular cross-section of 250 $\mu$m in the $y$-direction $\times$ 20 $\mu$m in the $z$-direction. It is then followed by several hyperbolic constrictions in series, all having the same inlet and outlet widths (200 and 20 $\mu$m respectively). In practice, RBC clusters go through the first two constrictions (1000 and 500 $\mu$m long) that allow to reach a near perfect centering and provide the initial (reference) state for the dissociation study that is performed in the third (250 $\mu$m  long) constriction. 

\begin{figure}[ht!]
\centering
 \includegraphics[width = \linewidth]{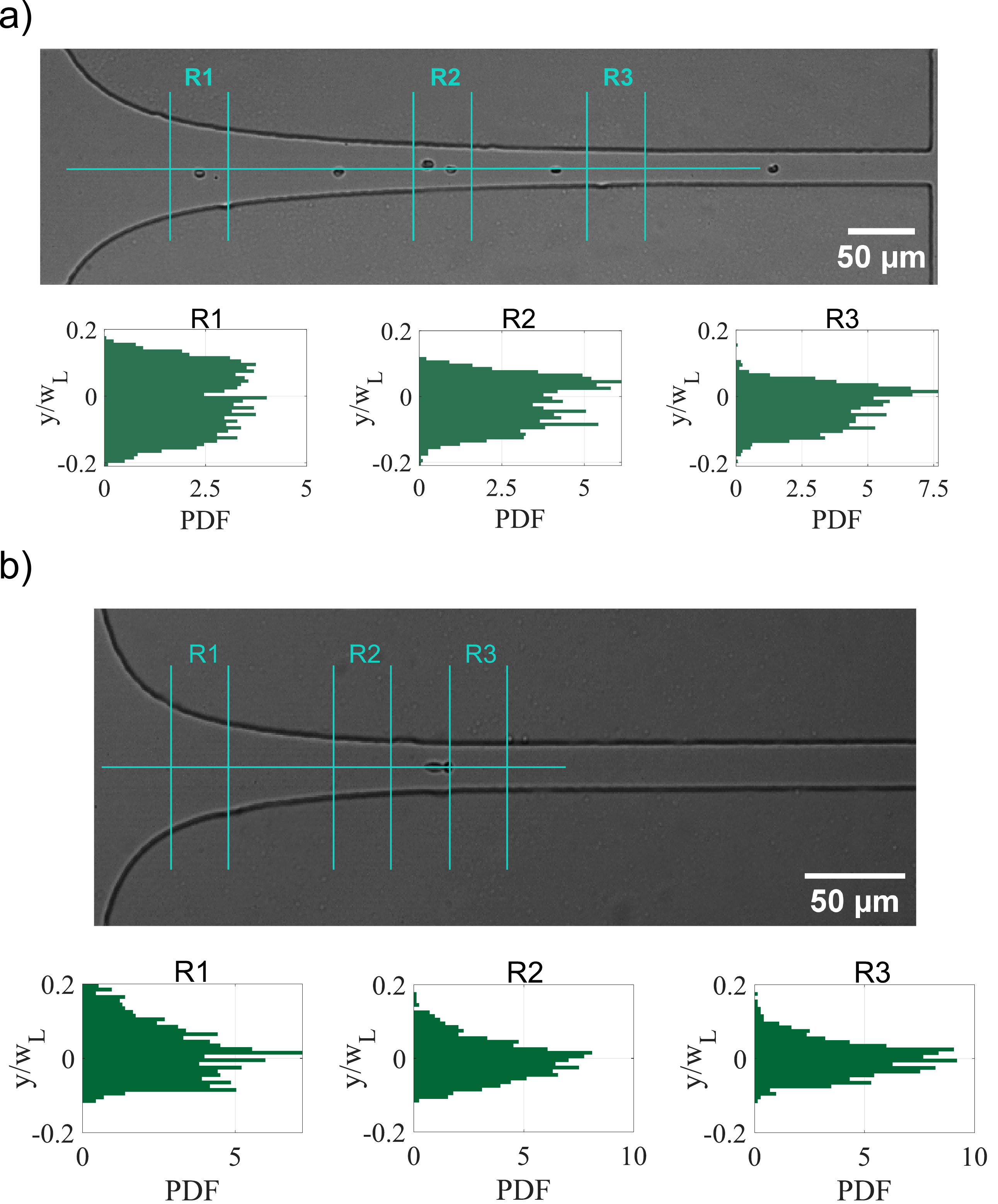}
  \includegraphics[width = \linewidth]{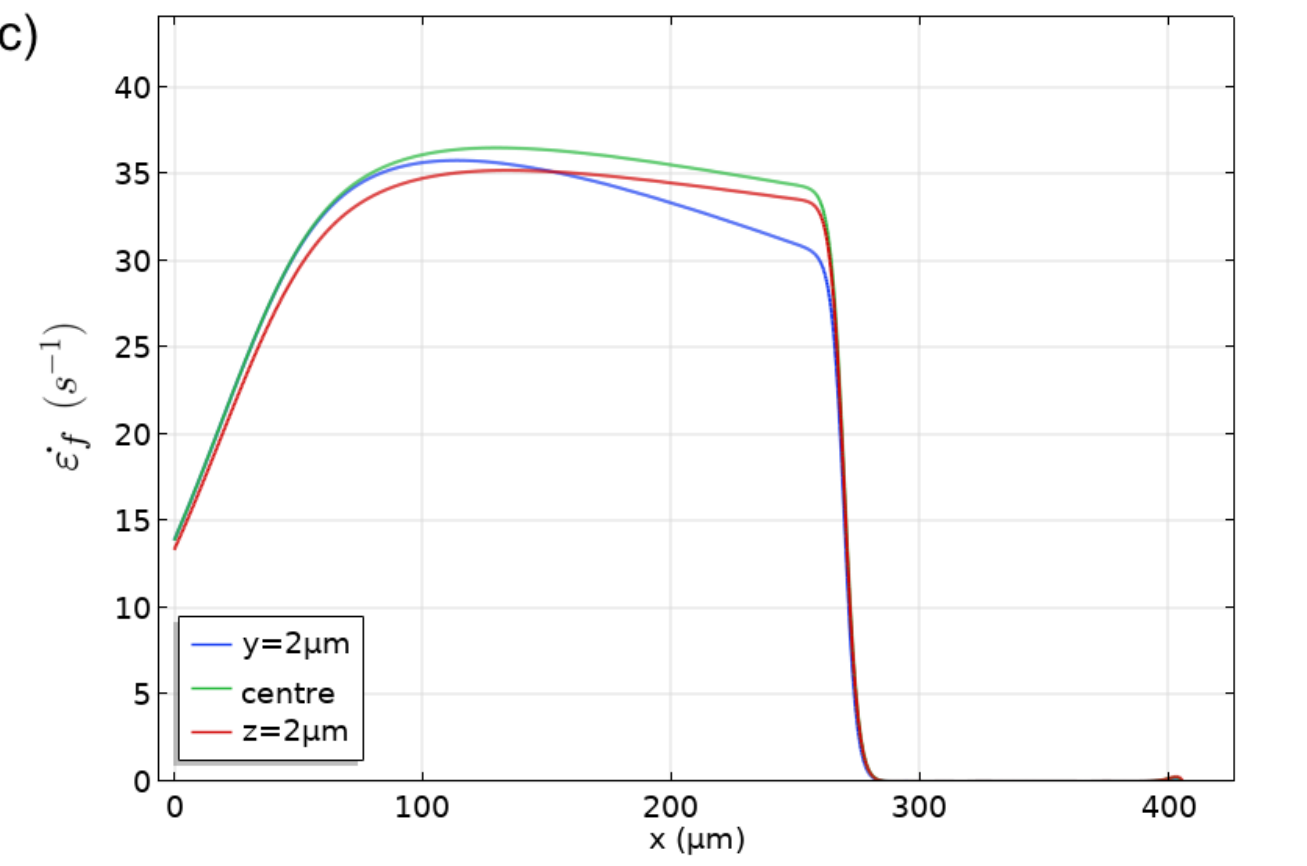}
  \caption{{Evaluation of the centering of cells in the $y$ direction in the 500 $\mu$m constriction (a) and 250 $\mu$m constriction (b). Top panels: Location of regions of interest R1, R2, R3 ((a): $x\in [88,118],~[293,323],~[440,470]~\mu$m respectively; (b) $x\in [56,86],~[136,166],~[194,224]~\mu$m respectively); Bottom panels: probability density function of the dimensionless coordinate $y/w_L$ showing that deviations from the centerline are small compared to the final width $w_L$ and in a range of $y$ coordinates where the flow is nearly invariant in the $y$-direction at R1 and R2 where most disaggregation events take place. (c): Fluid axial extension rate $\dot{\varepsilon}_f$ along two off-centered lines at $y=2~\mu$m and $z=2~\mu$m in the 250 $\mu$m constriction, from the same COMSOL simulation result as in Fig. \ref{fig:constriction}.}
  \label{fig:focusing}} 
\end{figure}

In order to evaluate the accuracy of the centering of cells and confirm that all analyzed objects experience similar hydrodynamic stress history, we quantified the dispersion of RBCs and clusters in the $y$-direction as they flow through the 250 $\mu$m long constriction as well as the previous one (500 $\mu$m long). Results are shown in Fig. \ref{fig:focusing}: we measured the $y$-positions of the centers of mass of objects in a typical recording (several thousands objects) in 3 successive regions of interest R1, R2, R3 in each constriction corresponding roughly to $L/3$, $2L/3$ and the end of the constriction just before the straight section.

When rescaling the $y$ positions with the final width of the constriction $w_L$, the results show that the standard deviations of $y/w_L$ from the center line at R1, R2 and R3 are respectively  $\pm 9$\%, $\pm 7$\%  and $\pm 6$\% in the 500 $\mu$m constriction and only $\pm 7$\%, $\pm 5$\%  and $\pm 5$\% in the 250 $\mu$m long studied constriction. This corresponds to deviations of only $\pm 1.4$, $\pm 1.0$ and $\pm 1.0$~$\mu$m, which is close to image resolution. 

Considering that the mean RBC diameter is about 7-8 $\mu$m i.e. large compared to these deviations and that in addition most recorded dissociations take place in the first two thirds of the constriction length (i.e. somewhere between R1 and R2 - see example of Fig. 1 in the article), where the channel aspect ratio $w(x)/h$ is greater than 1 and the velocity field varies only slowly with the $y$ coordinate around $y=0$, the hydrodynamic (extensional) stresses experienced by cells can be considered constant and insensitive to small deviations in $y$ or $z$ as shown in Fig. \ref{fig:focusing}c.

\section{Convolutional Neural Network}

A carefully assembled dataset (produced from image acquisitions in relevant regions of interest (ROIs) defined in the main text of the article) was curated to facilitate the application of machine learning techniques for the classification of red blood cell (RBC) aggregates based on the number of cells contained within each aggregate.
The dataset encompasses six distinct classes labeled from 0 to 5, where class 0 corresponds to images with no cells, 1 to 4 represent the number of RBCs in the corresponding aggregates, and class 5 denotes aggregates with 5 cells or more. 
The initial dataset consists of approximately 250 meticulously labelled images of RBC aggregates for each class, capturing a wide range of morphological characteristics and variations within each category. To enhance the diversity and robustness of the dataset, comprehensive data augmentation techniques were employed, resulting in the generation of a final dataset containing 1000 images per class. 

\begin{figure}[t!]
    \centering
    \includegraphics[width = \linewidth]{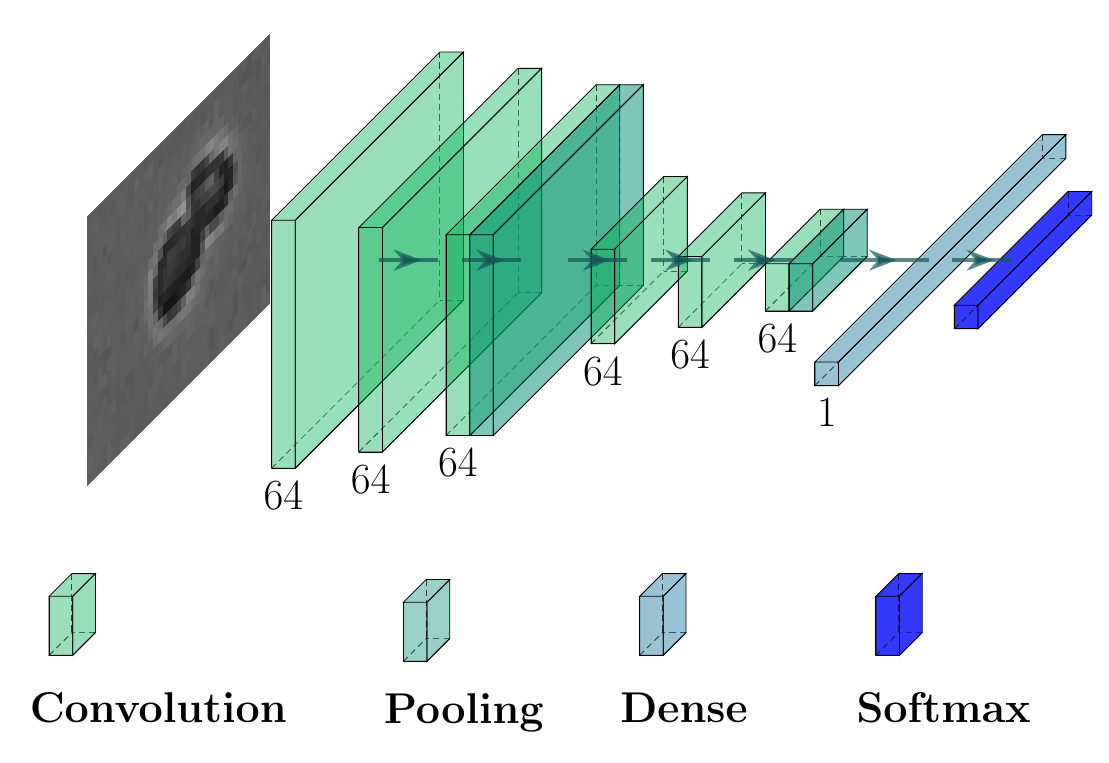}
    \caption{Architecture of the CNN used for classifying the aggregates based on the  number of cells contained within each.}
    \label{fig:CNN_arch}
\end{figure}

\begin{figure}[t!]
    \centering
    \includegraphics[width = \linewidth]{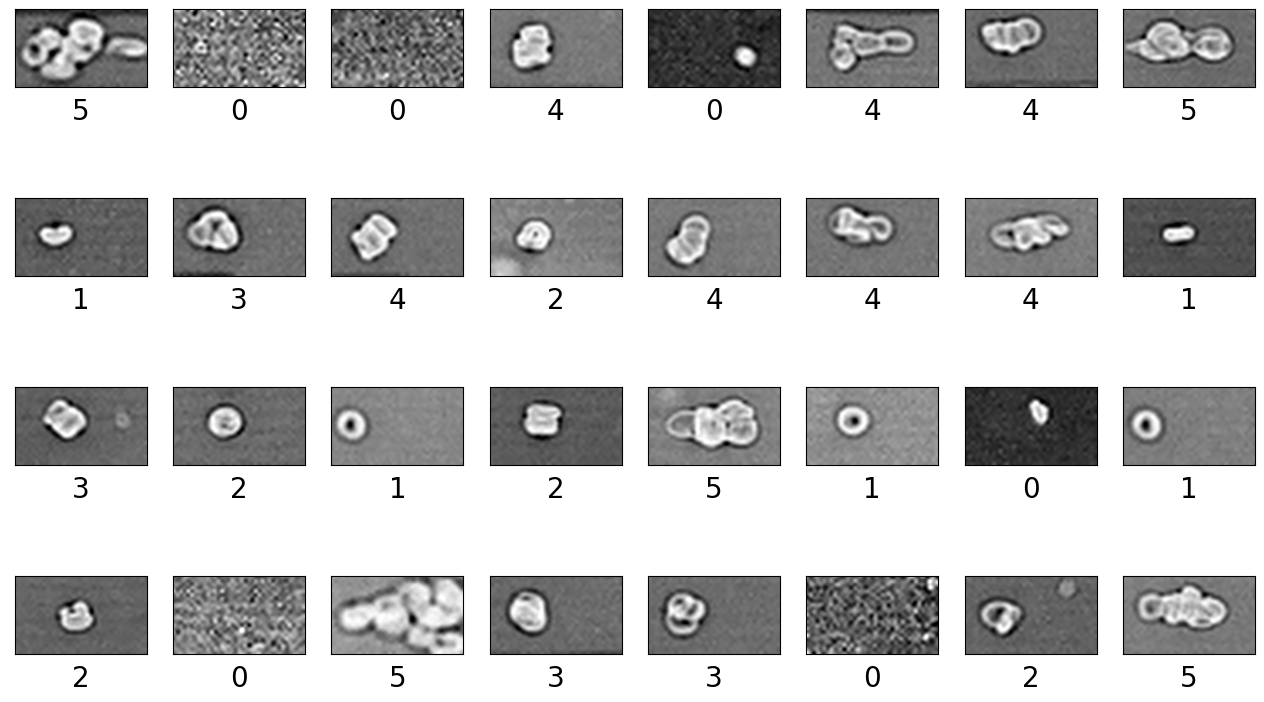}
    \caption{RBC aggregate images classified by the CNN where classes 0-4 correspond to the number of cells and class 5 is for aggregates of 5 cells or more.}
    \label{fig:CNN segmentation}
\end{figure}

A classification Convolutional Neural Network (CNN) model was then trained, utilizing the TensorFlow framework \cite{abadi2016tensorflow}.
The original dataset was divided into train, validation, and test sets, with proportions of 60\%, 20\%, and 20\%, respectively. 
To ensure a balanced representation of each class, the datasets were stratified, maintaining an equivalent number of samples from each class in each set. 

During the training process, the model used the sparse categorical cross-entropy loss function, which is well-suited for multi-class classification tasks. The Adam optimizer was employed, with a constant learning rate of $10^{-4}$ and a kernel of size $3\times3$ was used for all the convolution layers. The architecture of the CNN model is illustrated Fig. \ref{fig:CNN_arch}, showcasing the various layers and their connections. 

In the training phase, the model achieved an impressive accuracy of approximately 95\%, indicating its proficiency in accurately classifying RBC aggregates. Fig. \ref{fig:CNN segmentation} shows an example of categorized RBC aggregates alongside their corresponding class, ranging from class 0 (absence of RBCs in the image) to 5, (aggregates of five or more cells).
\\

\section{Variability between samples}

\begin{figure*}
     \resizebox{\textwidth}{!}{
   \includegraphics{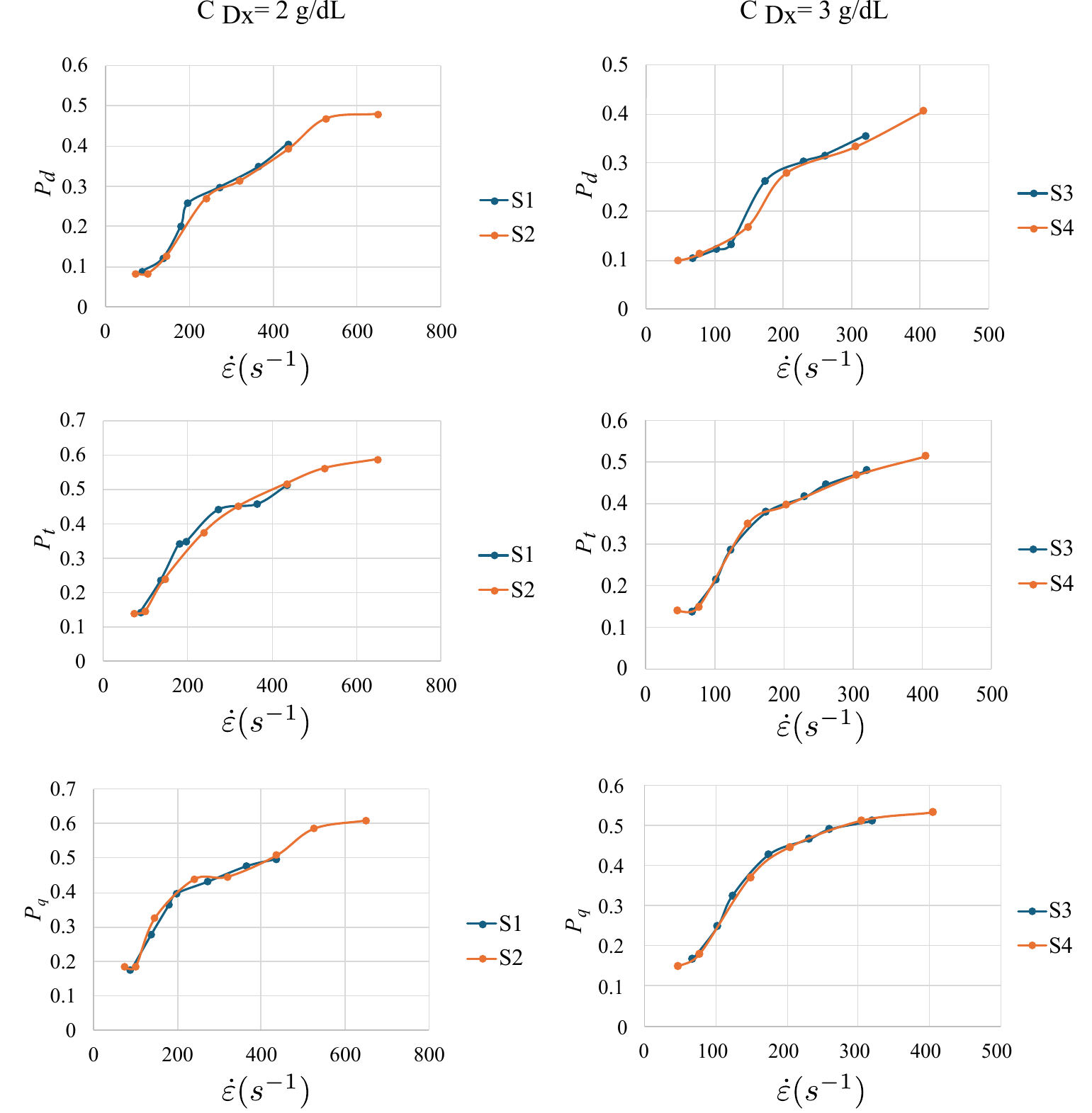}}
   \caption{{Dissociation probabilities $P_d$, $P_t$ and $P_q$ for Dextran 70 kDa concentrations of 2 g/dL (left column) and 3 g/dL (right column) where data for different samples are shown in different colors.}}
   \label{fig:donor_variability}
\end{figure*}

{A total of 7 samples were used in the presented results. For two values of the Dextran 70 kDa, 2 different samples were studied across the range of explored extension rates as a quick evaluation the variability between healthy donors. The results are shown in Fig. \ref{fig:donor_variability} where the same data as in Fig. 3 of the article is, represented with different colors. Remarkably, data from different donors (S1 and S2 for $c=2$~g/dL and S3 and S4 for $c=3$~g/dL fall on the same master curve with very little dispersion. While more samples would be needed for statistical and clinical relevance, this suggests that inter-individual variability of healthy RBC properties has a low impact on aggregation strength when cells are resuspended in the same buffer. Indeed, it is likely that inter-individual variability on the average properties of cells (size, hemoglobin content, membrane mechanical properties) is lower than the large intra-individual variations between cells (i.e. between cells of different age \cite{Waugh1992Mar,bosch1994,franco2013,Ermolinskiy2020}), which are already smoothed out by the statistical averaging provided by the data acquisition and processing procedure over thousands of aggregates.
}

\section{Estimation of lubrication effects in the separation process}

\REV{A close observation of the separation process (e.g. in Fig. 1 of the manuscript) reveals that it takes place in two steps: first a detachment or unbinding of membranes of two aggregated cells, during which the viscous extensional forces overcome adhesion forces until contact is lost, followed by the cells moving away from each other. \\
While the first stage is described by the force balance and scaling discussed in the manuscript, one may wonder whether viscous lubrication forces in the second phase could sufficiently slow down the subsequent separation process, so that the cells would not have moved far enough from each other before reaching the end of the constriction zone in the channel, thereby preventing them from being detected as separate cells and biasing the dissociation statistics. Here we estimate the timescale of this hydrodynamic separation process assuming a balance between lubrication forces between two separating spheres and viscous drag from the extensional flow.
In the lubrication approximation
\begin{equation}
F_{lub} \simeq -3 \pi \eta \frac{R^2}{\delta} \frac{d \delta}{dt}
\end{equation}
The viscous drag force acting on each sphere is given by equation 2 of the main manuscript:
\begin{equation}
F \sim  6 \pi \eta \dot{\varepsilon} R^2 
\end{equation}
where $\delta$ is the distance between particles.
The force balance $F_{lub}+F=0$ yields:
\begin{equation}
    2 \dot{\varepsilon} \simeq \frac{1}{ \delta} \frac{d \delta}{dt}
\end{equation}
Considering that the initial intermembrane distance is of order $\delta_i=20$ nm (see e.g. \cite{Chien80}) and that separation is clearly detectable by the image processing technique when the distance is $\delta_f=1$ µm, this will be achieved in a time $\tau_{sep}=ln(\delta_f/\delta_i) / (2 \dot{\epsilon}) = 9.6$~ms for an extension rate of 204 s$^{-1}$ as in the example of Fig. 1 where the timestep is 4~ms and the total transit time through the constriction of order 50~ms. \\
Therefore, the lubrication-mediated separation of cells, once they are dissociated is fast compared to the transit time and has a negligible impact on the detection of aggregate dissociation in the experiment.
\medskip
}

\bibliographystyle{unsrt}


\providecommand{\noopsort}[1]{}\providecommand{\singleletter}[1]{#1}%